\documentclass[aps,preprint,epsfig,rotate]{revtex4}
\usepackage{graphicx}
\usepackage{bm}
\usepackage{epsfig}

\begin{document}

\title{On bound state spectra of the one-electron diatomic ions} 

\author{Alexei M. Frolov}
 \email[E--mail address: ]{alex1975frol@gmail.com}  

\affiliation{Department of Applied Mathematics \\
 University of Western Ontario, London, Ontario N6H 5B7, Canada}

\date{November 27, 2026}

\begin{abstract}

The total energies of a large number of diatomic (or two-center) one-electron $A^{+} B^{+} 
e^{-}$ ions with unit electrical charges are determined numerically to high accuracy. Based 
on these results we derive some accurate mass-interpolation formulas for the total energies 
of such three-body systems (ions). These formulas can be applied to the both symmetric 
$A^{+} A^{+} e^{-}$ and non-symmetric $A^{+} B^{+} e^{-}$ diatomic ions. Based on the 
results obtained in this study we also consider a few actual and currently unsolved problems, 
which are known for the two-center (or diatomic) one-electron ions. \\  


\end{abstract}

\maketitle

\newpage
\section{Introduction}

In this communication we investigate the quasi-molecular (or two-center) one-electron ions with 
the unit electrical charges. These ions are designated below by the universal $A^{+} B^{+} 
e^{-}$ notation, where the point particles $A^{+}$ and $B^{+}$ are the two heavy, positively 
charged (bare) atomic nuclei, while $e^{-}$ is the negatively charged electron. The electric 
charge of each atomic nucleus equals $+ e$, while the electron has negative electric charge 
$- e$. A large number of similar systems, e.g., the H$^{+}_{2}$, HD$^{+}$, HT$^{+}$, D$^{+}_2$, 
DT$^{+}$ and T$^{+}_2$ ions are stable against any decay and they are easily observed in modern 
experiments. Similar three-body one-electron ions are of great interest in atomic and molecular 
physics, spectroscopy, astrophysics, plasma physics, solid state physics, physical chemistry 
and other areas of physics and chemistry. In this study we investigate the ground (bound) $1 s 
\sigma-$state energies of the three-body $A^{+} B^{+} e^{-}$ ions. Our main goal is to develop 
the new level of understanding of problems which exist in highly accurate computations of the 
bound states in three-, four- and few-body atomic systems which contain two heavy, positively 
charged particles.  In particular, we want to derive some accurate and simple mass-interpolation 
formulas for the total energies of the three-body $A^{+} B^{+} e^{-}$ ions. Another our aim is 
to re-evaluate the role of adiabatic (or Born-Oppenheimer) approximation \cite{BO} for highly 
accurate calculations of these two-center one-electron ions. In addition to this, we also discuss 
a number of new interesting problems which have been discovered in the course of our analysis.    

Currently, in the traditional atomic and molecular physics in order to perform accurate numerical 
calculations of bound states and describe the known experimental results, including many bound 
state properties of the $A^{+} B^{+} e^{-}$ ions, it is recommended to apply the adiabatic (or 
two-center) approximation \cite{BO} which is also called the Born-Oppenheimer approximation (see, 
e.g., \cite{Bethe} and \cite{LLQ}). Such an approximation is based on experimentally confirmed 
fact that the electron mass $m_e$ is significantly smaller than the mass of any atomic nucleus. 
In actual $A^{+} B^{+} e^{-}$ ions we have to describe and investigate one-electron motion in the 
field of two heavy, positively charged (Coulomb) nuclei. Here we have to note that there is only 
one pure adiabatic two-center system which is the model ${}^{\infty}$H$^{+}_{2}$ ion. This ion 
does not exist in reality, but it is extensively discussed in numerous theoretical studies. In 
the ${}^{\infty}$H$^{+}_{2}$ ion the both atomic nuclei have infinite atomic masses and they do 
not move et al during any electron motion. In actual $A^{+} B^{+} e^{-}$ ions each of the two 
nuclei can move and such a motion is affected by the electron motion and vice versa. It is clear 
that similar systems are quite different from the pure adiabatic model systems and systems close 
to them. The actual border between these two classes of $A^{+} B^{+} e^{-}$ ions is defined by 
the numerical values of nuclear masses (see below). 

As is well known the adiabatic approximation plays a central role in theoretical and experimental 
analysis of many actual molecular and quasi-molecular systems, solid state physics, plasma physics, 
etc. In many classical courses on solid state physics (see, e.g., \cite{Ziman} and \cite{Dav}) the 
adiabatic approximation is routinely introduced at very first pages. Davydov in \cite{Dav} generally 
wrote that the whole theory of a solid state could be impossible without adiabatic approximation. 
Adiabatic approximation is based on the fact that the ratio of electron mass $m_e$ to the masses of 
atomic nuclei $M$ (or $r = \frac{m_e}{M}$ ratio) is always less (and even substantially less) than 
$\approx 5 \cdot 10^{-4}$, i.e., it is a small number which is considered to be sufficient to apply 
the adiabatic approximation in the form of effective perturbation theory. 

However, as it was noticed in many papers, books and even textbooks (see, e.g., \cite{Bethe}, 
\cite{LLQ}, \cite{Eyr} - \cite{Herz}) that any computational method, which is applied for bound state 
computations and based on the adiabatic approximation, uses a different small parameter $\tau$ which 
equals $\tau = \sqrt[4]{\frac{m_e}{M}}$. This parameter is significantly larger and closer to unity 
than the mass ratio $r = \frac{m_e}{M}$ mentioned above. For instance, for the one-center hydrogen 
H$^{+}_{2}$, HD$^{+}$, HT$^{+}$ ions the parameter $\tau$ is $\approx \frac{1}{6.5}$, while the mass 
ratio $r \approx \frac{1}{2000}$. Briefly, this means that we can write the following chain of 
inequalities  $\tau > \frac{1}{10} \gg r \ge \frac{1}{2000}$. Since the parameter $\tau$ is not 
really small for many two-center systems, then we can predict a slow convergence rate for any 
numerical method which is based on the adiabatic approximation. This explains why all numerical 
procedures based on adiabatic approximation cannot produce any highly accurate results for relatively 
light two-center systems such as the H$^{+}_{2}$, HD$^{+}$, HT$^{+}$ ions and for other similar
systems which are often called the hydrogen `molecular' one-electron ions \cite{Eyr}. On the other 
hand, recently we were involved into fast and development of the general three-body methods and 
applications of these methods to various two-center, one-electron ions. For the last twenty years 
these methods have been transformed (see, e.g., \cite{Fro2002A} - \cite{Fro2003}) into a set of 
effective, compact and universal tools, which can be used for highly accurate variational 
calculations of arbitrary three-body systems, including two-center, one-electron ions. This 
essentially makes the adiabatic approximation useless for highly accurate computations of many 
actual two-center, one-electron ions which contain the nuclei of protium, deuterium and tritium. 

Such an unexpected and fundamental failure of the methods based on the adiabatic approximation 
opened a new road for the fast development and applications of different, pure three-body methods 
and approaches, which are new in this area of three-body systems and have no relation with the 
Born-Oppenheimer approximation. In particular, these new methods are of great interest for highly 
accurate calculations of bound states in the hydrogen `molecular' ions and systems close to them. 
Probably, one of the most successful and truly universal three-body methods has been developed in 
our papers \cite{Fro2002A} - \cite{Fro2003}, where we developed the basis set of complex exponents. 
In this set each of the basis function is written as the complex exponent of either relative 
($r_{32}, r_{31}, r_{21}$), or perimetric $(u_1, u_2, u_3)$ three-body coordinates \cite{Pek1}. 
These two basis sets are currently known as the exponential variational expansion(s) with the 
complex non-linear parameters \cite{Fro2002A}. In reality, by using our complex exponential 
variational expansions we could solve and  investigate many problems known for various 
one-electron two-center $A^{+} B^{+} e^{-}$ ions (see, e.g., \cite{Fro2018A} - \cite{Fro2021}). 
In general, none of these problems can be solved (even to relatively high accuracy) by using the 
methods based on the adiabatic (or Born-Oppenheimer) approximation. On the other hand, our 
complex variational expansion also works very well for the Ps$^{-}$ and H$^{-}$ ions, He-like 
atoms and ions and other three-body systems which have nothing to do with the adiabatic 
approximation. This was the main reason to call the complex variational expansion by the 
universal three-body variational expansion (see, e.g., our early papers \cite{Fro2002A}, 
\cite{Fro2002B}). 

This paper has the following structure. In the next Sections we consider the actual forms of 
Hamiltonian(s) which are used for bound state calculations and for theoretical analysis of the 
two-center one-electron ions. Construction of the three-body variational wave functions, which 
are used for highly accurate, bound state calculations of the $A^{+} B^{+} e^{-}$ ions, are 
discussed in Section III. Adiabatic divergence of the traditional variational expansions is 
described in the third and fourth Sections. Our computational results are discussed in Section 
V. Mass-interpolation formulas developed in this study are discussed and applied to actual 
systems in Section VI. The partial adiabatic limits for the  $A^{+} B^{+} e^{-}$ ions are 
briefly discussed in Section VII. Concluding remarks can be found in the last (eighth) Section. 
Vital relations between three-body perimetric coordinates and ellipsoidal coordinates are 
derived and explained in the Appendix A. In the Appendix B we discuss the completeness of our 
exponential basis set in the relative and/or perimetric coordinates. This problem is of 
paramount importance in various applications. In general, in this article, we want to focus on 
discussions of fundamental principles and ideas and delve less into technical details of our 
method. 

To conclude this Section, let us note that in our current analysis all nuclei are assumed to be 
point particles which have no internal structure. Also in this study we consider some atomic 
nuclei with very large and even infinite masses. It is clear that the $A^{+} A^{+} e^{-}$ and 
$A^{+} B^{+} e^{-}$ ions with such nuclei are the pure model systems. In reality, it is hard to 
imagine any atomic nucleus which has the unit electric charge, infinitely small radius (as a 
point particle) and mass larger than 10,000 $m_e$. However, in order to solve our problems 
formulated above we have to deal with similar (model) nuclei and molecular ions which contain 
such heavy nuclei. Therefore, the reader should not be surprised, if he encounters in the text 
below the notations such as $M = 500,000$ $m_e$, $M$ = 10,000,000 $m_e$, etc, and even $M = 
\infty$ $m_e$. Note that in contrast with earlier papers \cite{Fro2018A} - \cite{Fro2021}, our 
main attention below is devoted to the non-symmetrical $A^{+} B^{+} e^{-}$ ions which include 
one bound electron and two different heavy and very heavy atomic nuclei with the unit electric 
charges. 

\section{Hamiltonians used in bound state calculations} 

The non-relativistic Hamiltonian $\hat{H}$ of an arbitrary three-body, positively charged 
$A^{+} B^{+} e^{-}$ ion is written in the form 
\begin{eqnarray}
 \hat{H} = - \frac{\hbar^2}{2 M_{A}} \Delta_{1} - \frac{\hbar^2}{2 M_{B}} \Delta_{2} 
 - \frac{\hbar^2}{2 m_{e}} \Delta_{3} - \frac{Q_{B} e^2}{r_{32}} - 
 \frac{Q_{A} e^2}{r_{31}} + \frac{Q_A Q_B e^2}{r_{21}} \; , \; \label{Hamilt1}
\end{eqnarray}
where $M_{A}$ and $M_{B}$ are the nuclear masses of $A$ and $B$ nuclei, the operator 
$\Delta_{i} = \frac{\partial^{2}}{\partial x^{2}_{i}} + \frac{\partial^{2}}{\partial 
y^{2}_{i}} + \frac{\partial^{2}}{\partial z^{2}_{i}}$ is the Laplace operator of the 
particle $i$, while the potential energy is the sum of three Coulomb interactions 
between two point, electrically charged particles ($i$ and $j$), i.e., $\hat{V} = 
\sum_{(ij)} \frac{e^2}{r_{ij}} = \frac{e^2}{\mid {\bf r}_{i} - {\bf r}_{i} \mid}$, where 
$(ij) = (ji)$ = (32), (31), (21) and $(ij) = (ji)$. The electric charges of these two 
nuclei equal unity, i.e., we can assume below that $Q_{A} e = e$ and $Q_{B} e = e$. In 
this equation and everywhere below in this study the particles 1 and 2 mean the positively 
charged and heavy atomic (or hydrogen) nuclei $A$ and $B$, respectively, while the particle 
3 designates the electron $e^{-}$ which has the negative electric charge $- e$ and mass 
$m_{e}$. In this study we shall assume that the following condition (or constraint) for the 
three particle masses $m_e, M_A$ and $M_B$ is always obeyed: $M_B \ge M_A \gg m_e$. 

Also in Eq.(\ref{Hamilt1}) the notation $\hbar = \frac{h}{2 \pi}$ stands for the reduced Planck 
constant, which is often called the Dirac constant. In atomic units, where $\hbar = 1, e = 1$ 
and $m_e = 1$ (also $4 \pi \epsilon_0 = 1$), the same Hamiltonian $\hat{H}$, Eq.(\ref{Hamilt1}), 
takes the form 
\begin{eqnarray}
 \hat{H} = - \frac{1}{2 M_{A}} \Delta_{1} - \frac{1}{2 M_{B}} \Delta_{2} - \frac{1}{2} 
 \Delta_{3} - \frac{1}{r_{32}} - \frac{1}{r_{31}} + \frac{1}{r_{21}} \; , \; \label{Hamilt}
\end{eqnarray} 
where $M_{A} = \frac{M_A}{m_e} = \Bigl(\frac{M_A}{m_e}\Bigr) \; m_e$ and $M_{B} = \frac{M_B}{m_e} 
= \Bigl(\frac{M_B}{m_e}\Bigr) \; m_e$ in atomic units. In the two-center molecular $A^{+} B^{+} 
e^{-}$ ions the internuclear distance $r_{21}$ in Eq.(\ref{Hamilt}) is often designated as $R$. 
The Hamiltonian, Eq.(\ref{Hamilt}), includes the electron's motion and all possible motions of 
two nuclear particles, e.g., their rotations and `vibrations'. The Hamiltonian, 
Eq.(\ref{Hamilt}), has been extensively used in many papers, including this study, to perform 
highly accurate numerical computations of different two-center ions. It is clear that such 
calculations can be done only in such sets of basis functions which are rotationally and 
translationally invariant. In reality, our exponential basis sets (see, e.g., Eqs.(\ref{expc}) 
and (\ref{expc1}) below) are rotationally and translationally invariant. Indeed, each basis 
functions from these two sets has such an invariance. Therefore, the linear combinations of 
these basis functions will be also rotationally and translationally invariant. 

The explicit forms of the both Hamiltonians, Eq.(\ref{Hamilt1}) and Eq.(\ref{Hamilt}), are very 
general, but they are not convenient for analytical investigations of inter-nuclear motions in real 
$A^{+} B^{+} e^{-}$ ions. In such cases the original Hamiltonians, Eq.(\ref{Hamilt1}) and 
Eq.(\ref{Hamilt}), are usually re-written in a few different forms. For instance, the Hamiltonian of 
an arbitrary three-body $A^{+} B^{+} e^{-}$ ion can be represented in the following approximate form 
which is convenient to conduct analytical investigations of the internal motions, i.e., motions 
inside of this molecular ion:  
\begin{eqnarray}
 \tilde{H} = &-& \frac{\hbar^{2}}{2 \; m_e} \Bigl(\frac{\partial^{2}}{\partial x^{2}_{3}} 
 + \frac{\partial^{2}}{\partial y^{2}_{3}} + \frac{\partial^{2}}{\partial z^{2}_{3}}\Bigr) 
 - \frac{\hbar^{2}}{2 M R^{2}} \Bigl[ \frac{\partial}{\partial R} \Bigl(R^{2} 
 \frac{\partial}{\partial R}\Bigl) \; + \; \frac{1}{\sin \theta} \; 
 \frac{\partial}{\partial \theta} \Bigl( \sin \theta \; \frac{\partial}{\partial 
 \theta}\Bigr) \nonumber \\
 &+& \frac{1}{\sin^{2} \theta} \; \frac{\partial^{2}}{\partial \phi^{2}} \Bigr] + 
 V( x_{3}, y_{3}, z_{3}; R, \theta, \phi) \; \; , \; \label{Hamiltap}
\end{eqnarray} 
where $R = \mid {\bf R}_2 - {\bf R}_1 \mid = r_{21}$ is the scalar distance between two heavy nuclei 
which have masses $M_A$ and $M_B$, while $M$ is the reduced nuclear mass, i.e., $M = \frac{M_A \; 
M_B}{M_A + M_B}$. The origin of coordinates is located in the middle of internuclear distance $R$, 
while the molecular axis (or with the $A^{+} - B^{+}$ line in our case) coincides with the $z-$axis. 
In addition to this, the lighter $A-$nucleus has a positive $z-$coordinate and heavier $B-$nucleus 
has a negative $z-$coordinate. In this case, the two angles $\theta$ and $\phi$ in 
Eq.(\ref{Hamiltap}) are the polar and azimuthal angles of the electron in this molecular ion. Here 
we assume that the notations $x_3, y_3$ and $z_3$ stand for the Cartesian coordinates of the third 
particle (electron). Then, the two angles $\theta$ and $\phi$ in Eq.(\ref{Hamiltap}) are the polar 
and azimuthal angles of the electron in this molecular $A^{+} B^{+} e^{-}$ ion. 

This approximate Hamiltonian, Eq.(\ref{Hamiltap}), is used for analytical investigation of the 
internal motions in the two-center molecular ion, e.g., nuclear vibrations, rotations and their 
interactions with each other and with the electron motion. These problems are not considered in 
this study. Therefore, below the Hamiltonian, Eq.(\ref{Hamiltap}), is not used. Another (third) 
form of the Hamiltonians, Eq.(\ref{Hamilt1}) and Eq.(\ref{Hamilt}), is discussed below in 
Section IV. 

\section{Variational functions for bound state calculations} 

In general, to determine the total energy of an arbitrary bound state in three-body systems we
need to solve the corresponding Schr\"{o}dinger equation $\hat{H} \Psi = E \; \Psi$, where $E$ 
is the total energy of this state and $\Psi$ is its wave function. Fortunately, we can avoid  
this difficult task by replacing the original problem by the minimization of energy functional 
$E$: 
\begin{eqnarray}
  E = \min_{\Psi} \; \frac{\langle \Psi \mid \hat{H} \mid \Psi \rangle}{\langle \Psi \mid \Psi 
  \rangle} = \min_{\Psi_{N}} \; \langle \Psi_N \mid \hat{H} \mid \Psi_N \rangle \; , 
  \; \label{prop1}
\end{eqnarray}
where $\hat{H}$ is the Hamiltonian, while $E$ is its eigenvalue and $\Psi$ is the corresponding  
eigenfunction which has the finite norm. In Eq.(\ref{prop1}) the notation $\Psi_N$ stands for 
the corresponding unit-norm wave function $\Psi_N = \frac{\Psi}{\sqrt{\langle \Psi \mid \Psi 
\rangle}}$. As is well known (see, e.g., \cite{LLQ}) such a unit-norm wave function can be 
defined for an arbitrary bound state. Everywhere below in this study we shall designate the 
bound state wave functions by the notation $\Psi$, while in reality they are the $\Psi_{N}$ 
functions. 

Now, let us discuss the two different forms of our exponential three-body variational expansions   
which are extensively used in this study. These expansions are written in one of the two 
following forms (see, e.g., \cite{Fro2002A} and \cite{Fro2002B} and earlier references therein):
\begin{eqnarray}
 \Psi_{LM} = \frac{1}{2} (1 + \kappa \hat{P}_{21}) \sum_{i=1}^{N} \sum^{L}_{\ell_{1}=0} 
 C_{i}(\ell_{1}(i), \ell_{2}(i); L) \; {\cal Y}_{LM}^{(\ell_{1},\ell_{2})} ({\bf r}_{31}, 
 {\bf r}_{32}) \exp(-\alpha_{i} r_{32} - \beta_{i} r_{31} - \gamma_{i} r_{21}) \; \; 
 \; , \; \; \label{expc} 
\end{eqnarray}
and 
\begin{eqnarray}
 \Psi_{LM} = \frac{1}{2} (1 + \kappa \hat{P}_{21}) \sum_{i=1}^{N} \sum^{L}_{\ell_{1}=0} 
 C_{i}(\ell_{1}(i), \ell_{2}(i); L) \; {\cal Y}_{LM}^{(\ell_{1},\ell_{2})} ({\bf r}_{31}, 
 {\bf r}_{32}) \exp(-\alpha_{i} u_1 - \beta_{i} u_2 - \gamma_{i} u_3) \; \; \; , 
 \; \; \label{expc1} 
\end{eqnarray}
where the first series is called the three-body exponential variational expansion in the 
relative coordinates $r_{32}, r_{31}, r_{21}$, while the second expansion is the three-body 
exponential variational expansion in the perimetric coordinates $u_{1}, u_{2}, u_{3}$. The 
both relative and perimetric coordinates are closely related to each other (see Appendix A 
below). It is also clear that the both these basis sets are translationally and rotationally 
invariant. The operator $\hat{P}_{21}$ describes permutations of the two identical nuclei in 
the $A^{+} A^{+} e^{-}$ ion, when the parameter $\kappa = \pm 1$. For the non-symmetric 
$A^{+} B^{+} e^{-}$ ions  we have to assume that the parameter $\kappa$ in Eq.(\ref{expc1}) 
equals zero. The angular functions ${\cal Y}_{LM}^{\ell_{1},\ell_{2}} ({\bf r}_{31}, 
{\bf r}_{32})$ are the bipolar harmonics defined exactly as in \cite{Varsh} (see also 
\cite{Fro2002B}). In general, it is also necessary to prove the completeness of each set of 
basis functions used to construct our trial wave functions in Eq.(\ref{expc}) and 
Eq.(\ref{expc1}) above. This problem is of great interest in numerous applications of the 
exponential variational expansions and it is considered in the Appendix B below.  

The non-linear parameters $\alpha_{i}, \beta_{i}, \gamma_{i}$ (for $i = 1, \ldots, N$) in 
Eqs.(\ref{expc}) and (\ref{expc1}) are assumed to be complex and their imaginary parts cannot 
be equal zero all at once. For simplicity, below we shall consider only bound states with $L 
= 0$ (and $M = 0$). In this case our variational expansions, Eq.(\ref{expc}) and 
Eq.(\ref{expc1}), take the forms  
\begin{eqnarray}
 \Psi_{0 0} = \frac{1}{2} (1 + \kappa \hat{P}_{21}) \sum_{k=1}^{N} C_{k} \exp(-\alpha_{k} 
 \; r_{32} - \beta_{k} \; r_{31} - \gamma_{k} \; r_{21} - \imath \; d_{k} \; r_{32} - 
 \imath \; e_{k} \; r_{31} - \imath \; f_{k} \; r_{21}) \; \; \; , \; \; \label{expc2} 
\end{eqnarray}
and 
\begin{eqnarray}
 \Psi_{0 0} = \frac{1}{2} (1 + \kappa \hat{P}_{21}) \sum_{k=1}^{N} C_{k} 
 \exp(-\alpha_{k} \; u_1 - \beta_{k} \; u_2 - \gamma_{k} \; u_3 - \imath \; d_{k} \; 
 u_1 - \imath \; e_{k} \; u_2 - \imath \; f_{k} \; u_3) \; \; \; , \; \; \label{expc3} 
\end{eqnarray}
respectively. In these two formulas all $6 N$ non-linear parameters $\alpha_{k}, \beta_{k}, 
\gamma_{k}, d_{k}, e_{k}, f_{k}$, where $i = 1, 2 \ldots, N$, are real, while the notation 
$\imath$ stands for the imaginary unit, i.e., $\imath^{2} = - 1$. It is clear that in 
Eq.(\ref{expc2}) and Eq.(\ref{expc3}) all non-linear parameters, which are designated by 
the Greek letters $\alpha_{k}, \beta_{k}, \gamma_{k}$ must always be positive. These 
conditions are needed to guarantee convergence for all essential three-body integrals. 
Optimization of these non-linear parameters allows one to produce highly accurate and even 
extremely accurate variational wave functions for arbitrary, in principle, two-center, 
one-electron $A^{+} B^{+} e^{-}$ ions. On the other hand, by performing simple numerical 
experiments with Eq.(\ref{expc2}) and Eq.(\ref{expc3}) one easily finds that when the total 
number of such parameters is large (or even very large), then any detailed optimization of 
each single non-linear parameter becomes substantially less important for the overall 
accuracy of the variational result. However, in such cases it is crucially important to  
increase the total number of the number of basis wave function in Eq.(\ref{expc2}) and 
Eq.(\ref{expc3}). In particular, for the both symmetric $A^{+} A^{+} e^{-}$ and 
non-symmetric $A^{+} B^{+} e^{-}$ diatomic ions this method works very well and we used 
it in this study. 

Our computational results are the total energies (in $a.u.$) of a large number of the 
model symmetric $A^{+} A^{+} e^{-}$ and non-symmetric $A^{+} B^{+} e^{-}$ ions which 
can be found in Tables I and II. Table II also contains important details about overall 
convergence rates of our results (i.e., total energies) obtained for some selected 
$A^{+} B^{+} e^{-}$ ions. The ions from Table II were selected almost randomly, since 
results for other similar ions shown in Table I also converge fast. Some details of our 
variational calculations are shown in Table II. The data from Table II explains how we 
evaluate the total number of stable decimal digits for each bound $A^{+} B^{+} e^{-}$ 
ion. Note that all our calculations in this study have been performed by using the 
extended arithmetic precision which is provided by the Fortran pre-translator written 
by David H. Bailey \cite{Bail1}, \cite{Bail2}. This allowed us to achieve the overall 
numerical accuracy $\approx$ 64 - 84 - 116 exact decimal digits. 

The functions, Eqs.(\ref{expc}) - (\ref{expc3}), are the three-body variational expansions 
which can be applied to arbitrary, in principle, three-body systems. Our computational 
results obtained for various bound states in different three-body systems indicate clearly 
that the exponential variational expansion(s), Eqs.(\ref{expc}) - (\ref{expc3}), provide 
highly accurate approximations for the actual three-body wave functions. However, in our 
earlier computational studies of three-body systems we could apply the `classical' 
exponential variational expansions, Eq.(\ref{expc1}), with the real non-linear parameters 
only. This led to a significant slowdown in actual convergence rates for all molecular 
$A^{+} B^{+} e^{-}$ ions. Furthermore, this effect rapidly did increase when the masses of 
two heavy`nuclei' grow. Originally, this effect was found in the series of ground (bound) 
states in the muonic molecular ions $p p \mu, d d \mu$ and $t t \mu$. In general, the total 
energies $E(N)$ converge to some `final' (or `exact') $E(N = \infty) = E(\infty)$ values 
and this process is well described by the following asymptotic formula(s): 
\begin{eqnarray}
  E(N) = E(\infty) + \frac{A}{N^{\gamma}} \; \; \; , \; \; \; {\rm or} \; \; \; 
  E(N) = E(\infty) + \frac{A_1}{N^{\gamma}} + \frac{A_2}{N^{\gamma + 1}} \; \; \; . \; \; 
  \label{conv} 
\end{eqnarray}

After a few applications to actual three-body systems one easily finds that the numerical 
parameter $\gamma$ plays a crucial role here. The larger parameter $\gamma$ in this 
formula means the faster convergence rate. For the ground (bound) states in the muonic 
molecular ions $p p \mu, d d \mu$ and $t t \mu$ we have found that the parameter $\gamma$ 
in Eq.(\ref{conv}) decreases from 8 - 8.5 for the $p p \mu$ ion to 5 - 5.5 for the $t t 
\mu$ ion (for the $d d \mu$ ion this parameter equals $\approx$ 6.5). Thus, the overall 
convergence rate of the `classic' exponential expansion is slowing down (and even 
substantially), if we move from the $p p \mu$ ion to the $t t \mu$ ion. This phenomenon 
was also observed for other three-body systems and for different variational expansions 
known at that time for three-body systems. In particular, we have found (see, e.g., 
\cite{Fro1987}) that for the ${}^{1}$H$_{2}^{+}$ ion this crucial parameters $\gamma$ was 
less than 2. This means that our variational expansion, Eq.(\ref{expc1}), with the real 
non-linear parameters only cannot be used, ib principle, for accurate bound state 
calculations of this simple three-body ion. Such a phenomenon was called the `adiabatic', 
or `two-center' divergence of our three-body exponential variational expansion 
Eq.(\ref{expc1}) (see, e.g., \cite{Fro1987}). Further investigations indicate clearly 
that the same problem does exist for other three-body variational expansions which is 
written in the relative coordinates $r_{32}, r_{31}, r_{21}$, e.g., for the Hylleraas 
and/or semi-exponential \cite{Fro2010} variational expansions. This problem is quite 
common for atomic three-body one-electron systems with the two heavy positively charged 
nuclei. Therefore, it is absolutely necessary to understand the fundamental reasons of 
this interesting phenomenon. This is our goal in the next Section.  

\section{Adiabatic divergence in one-electron two-center ions} 

Slow convergence of the `traditional' three-body variational expansions for the two-center 
one-electron systems can be explained by the fact that any of these `traditional' 
expansions cannot describe well, i.e., to high accuracy, the new internal `nuclear' motions 
which arise in such systems. First of all, there are some nuclear vibrations of heavy nuclei 
which are relatively slow in time (in comparison with the electron's motion), but play an 
important role in the two-center one-electron ions. Second, for any two-center one-electron 
systems there is a problem of inter-nuclear localization. Briefly, this means that in any 
$A^{+} B^{+} e^{-}$ ion we know the actual distance between two heavy, positively charged 
particles (nuclei) $a$ $priory$, i.e., even before our calculations started. In the result 
of our calculations we need to find that $r_{21} = R$, where $R$ is some positive real 
number real number which can be predicted before numerical calculations. This is called 
the inter-nuclear localization, or nuclear localization, for short. It is clear that the 
classical variational expansion, Eq.(\ref{expc1}), with the real non-linear parameters 
cannot describe nether nuclear vibrations, nor actual nuclear localization in the heavy 
two-center $A^{+} B^{+} e^{-}$  ions. 

The adiabatic divergence can be understood from the first principles of the theory, if we 
consider the Hamiltonian of the problem. Unfortunately, we cannot deal with the Hamiltonian 
written in the form, Eq.(\ref{Hamilt}), since it contains some other non-essential motions, 
e.g., translations and rotations, which have nothing to do with the internal motions in the 
three-body $(A B e)^{+}$ ion. The corresponding Hamiltonian which is appropriate for our 
purposes here is written in the relative coordinates $r_{32}, r_{31}, r_{21}$. For the sake 
of simplicity, here we shall discuss only the bound states in these systems with $L = 0$. 
The non-relativistic Hamiltonian $H$ of the $(A B e)^{+}$ system (or ion) in the relative 
coordinates and in atomic units takes the form 
\begin{eqnarray}
 H = &-& \Bigl(\frac{1}{2 m_e} + \frac{1}{2 M_B}\Bigr) \Bigl[ \frac{\partial^2}{\partial 
  r^{2}_{32}} + \frac{2}{r_{32}} \frac{\partial}{\partial r_{32}} \Bigr] - \frac{1}{2 
  m_e} \Bigl[\frac{r^{2}_{32} + r^{2}_{31} - r^{2}_{21}}{r_{32} r_{31}}\Bigr] 
  \frac{\partial^2}{\partial r_{32} \partial r_{31}} \nonumber \\
  &-& \Bigl(\frac{1}{2 m_e} + \frac{1}{2 M_A}\Bigr) \Bigl[ \frac{\partial^2}{\partial 
   r^{2}_{31}} + \frac{2}{r_{31}} \frac{\partial}{\partial r_{31}} \Bigr] - 
   \frac{1}{2 M_B} \Bigl[\frac{r^{2}_{32} + r^{2}_{21} - r^{2}_{31}}{r_{32} r_{21}}\Bigr] 
  \frac{\partial^2}{\partial r_{32} \partial r_{21}} \nonumber \\
  &-& \Bigl(\frac{1}{2 M_A} + \frac{1}{2 M_B}\Bigr) \Bigl[ \frac{\partial^2}{\partial 
   r^{2}_{21}} + \frac{2}{r_{21}} \frac{\partial}{\partial r_{21}} \Bigr] - 
   \frac{1}{2 M_A} \Bigl[\frac{r^{2}_{31} + r^{2}_{21} - r^{2}_{32}}{r_{31} 
   r_{21}}\Bigr] \frac{\partial^2}{\partial r_{31} \partial r_{21}} \nonumber \\
   &-& \frac{1}{r_{32}} - \frac{1}{r_{31}} + \frac{1}{r_{21}} \; \; \; \label{Hamil}
\end{eqnarray}
where the particles 1 and 2 are the two heavy nuclei, e.g., the nuclei of two hydrogen isotopes, 
while particle 3 is the electron ($e^{-}$). As follows from Eq.(\ref{Hamil}) in the limit when 
$M_A \rightarrow \infty$ and $M_B \rightarrow \infty$ (or $\min (M_A, M_B) \rightarrow \infty$) 
all terms which include derivatives in respect to the $r_{21}$ variable, i.e., the  
$\frac{\partial}{\partial r_{21}}$ and/or $\frac{\partial^2}{\partial r^{2}_{21}}$ terms, vanish 
from the Hamiltonian, Eq.(\ref{Hamil}). This means that in the adiabatic limit, i.e. when $\min 
(M_A, M_B) \rightarrow \infty$, the internuclear variable $r_{21} = R$ becomes an additional 
parameter of the three-body problem,  since it does not change during any electron's motion. In 
other words, the internuclear variable $r_{21} = R$ is a constant and cannot be considered as an 
actual coordinate for this problem. 

The fact that the variable $r_{21}(= R)$ is a constant during the electron's motion in the pure
adiabatic (or two-center) systems means that the `exact' wave function $\Psi(r_{32}, r_{31}, 
r_{21})$ of the truly adiabatic system is represented in the form 
\begin{eqnarray}
   \Psi(r_{32}, r_{31}, r_{21}) = \delta(r_{21} - R) \phi(r_{32}, r_{31}) \; \; \; \label{phi}
\end{eqnarray}
where $\phi(r_{32}, r_{31})$ is the electron-nuclear part (or regular part) of the total 
three-particle wave function $\Psi$. In reality, the explicit form of the `exact wave function, 
Eq.(\ref{phi}), is appropriate only for the truly adiabatic two-center systems, e.g., for the 
${}^{\infty}$H$^{+}_{2}$ ion and close systems. For actual three-body systems which include two 
heavy (but not infinitely heavy!) atomic nuclei and one bound electron the exact delta-function 
in Eq.(\ref{phi}) must be replaced by some $\delta-$like spatial distribution, e.g., by the 
following one-parametric function 
\begin{eqnarray}
   \delta(r_{21} - R) \approx \frac{A}{\pi [A^2 + (r_{21} - R)^2]} \; \; \; \label{delta}
\end{eqnarray}
where $A$ is a numerical parameter. In the limit $A \rightarrow 0$ this function approaches to 
the actual two-center delta-function. This part of adiabatic divergence is directly related with 
the `localization' of two very heavy particles (atomic nuclei) at some certain internuclear 
distance. Usually, the overall contribution of the inter-nuclear `localization' is relatively 
large only for pure adiabatic ions, e.g., for those $A^{+} B^{+} e^{-}$ ions where $M_A \gg 1 
\cdot 10^{8}$ $m_e$ and $M_B (\ge M_A) \gg 1 \cdot 10^{8}$ $m_e$.  

Another source of the adiabatic divergence is directly related with the `vibrational' motion of 
the two heavy atomic nuclei in the non-symmetric $A^{+} B^{+} e^{-}$ and symmetric $A^{+} A^{+} 
e^{-}$ ions. The relative role of such `vibrations' is slightly larger for the three-body $A^{+} 
B^{+} e^{-}$ ions with lighter atomic nuclei. In general, any oscillating (or vibrational) motion 
of the two heavy particles in three-particle systems cannot be described accurately, if one 
applies the traditional exponential variational expansion, Eq.(\ref{expc1}), with the real and 
positive non-linear parameters only. In fact, if we apply any real exponential expansion, 
Eq.(\ref{expc1}), to the two-center systems, then the power parameter $\gamma$ in Eq.(\ref{conv}) 
rapidly fails to some small values ($\le 4$) and such an expansion cannot be considered as highly 
accurate even in the first approximation. In contrast with this our modified exponential 
expansions, Eq.(\ref{expc2}) and Eq.(\ref{expc3}), with the complex non-linear parameters always 
provide very high convergence rate for any three-body systems, including the two-center adiabatic 
systems. In reality, this follows from numerous results of actual computations with our advanced 
exponential variational expansions, Eq.(\ref{expc2}) and Eq.(\ref{expc3}), which contain the 
complex non-linear parameters. In other words, these variational expansions form the appropriate, 
rapidly convergent and highly-accurate tool for complete solution of the problems of nuclear 
`locations' and `vibrations' in any one-electron, two-center ion and this is the final solution 
of the `adiabatic divergence' which were known for years in actual atomic calculations. In our 
earlier papers the both exponential variational expansions, Eq.(\ref{expc2}) and 
Eq.(\ref{expc3}), were called the universal expansion for non-relativistic three-body systems. 
This means that for an arbitrary such system these two variational expansions can quickly produce 
highly accurate variational results.  

\section{Mass-interpolation formulas for the total energies} 

The total energies and many other bound state properties of the symmetric $A^{+} A^{+} e^{-}$ 
and non-symmetric $A^{+} B^{+} e^{-}$ ions are quite unique, since they can be approximated 
to very good numerical accuracy by using some general mass-interpolation (or mass-asymptotic) 
formulas constructed for each of these properties. In reality, this works in the following way. 
First, we select a number of `anchor', or `fulcrum' three-body $A^{+} B^{+} e^{-}$ ions with 
different nuclear masses and determine the total energies and other bound state properties of 
these ions. Then, by using these energies we have to develop some accurate mass-interpolation 
formulas for the energies and bound state properties. The principal moment here is a deep 
understanding of all the features of mass-dependence of the total energy in an arbitrary 
three-body system. In other words, it is absolutely necessary to know the exact analytical 
expression of the energy $E( m_1, m_2, m_3)$ function upon the masses of particles. Otherwise, 
such a mass-interpolation formula cannot be accurate. If this condition is obeyed, then at the 
final step these formulas can be used for highly accurate and even analytical predictions of 
the total energies and other bound state properties of the new $A^{+} B^{+} e^{-}$ ions without 
performing any additional calculations. 

In order to construct any accurate mass mass-interpolation formulas for the bound states in the 
$A^{+} B^{+} e^{-}$ ions one needs to perform a significant amount of highly accurate numerical 
computations of these ions. In addition to this, we have to come up with some `internally 
correct' mass-interpolation (or mass-asymptotic) formula or the total energies of the two-center, 
one-electron ions with unit electrical charges, i.e., for the non-symmetric $A^{+} B^{+} e^{-}$ 
and symmetric $A^{+} A^{+} e^{-}$ ions. In general, the total energy of an arbitrary three-body 
system with the three given masses $m_1, m_2, m_3$ and three unit electrical charges is 
represented in the following three-parameter form 
\begin{eqnarray}
   E = \frac12 \; Ry \; \alpha \; \Phi( \beta, \gamma) = \frac{m_e \; e^{4}}{2 \; \hbar^{2}} \; 
   \alpha \; \Phi( \beta, \gamma) \; \; , \; \label{mass1}
\end{eqnarray}
where $Ry = \frac{m_e \; e^{4}}{\hbar^{2}}$ is the atomic Rydberg constant, while the dimensionless 
parameters $\alpha, \beta$ and $\gamma$ are: 
\begin{eqnarray}
  \alpha = \frac{m_1 \; m_2 \; m_3}{m_e \; D} \; \; , \; \; \beta = \frac{m_3 ( m_1 + m_2 )}{2 
  \; D} \; \; , \; \; \gamma = \frac{m_3 ( m_2 - m_1 )}{2 \; D} \; \; , \; \label{mass2}
\end{eqnarray}
where $D = m_1 \; m_2 + m_1 \; m_3 + m_2 \; m_3$. These variables (or parameters) can be applied 
to arbitrary three-body systems, including Coulomb three-body systems and two-center one-electron 
ions. Formally, in Eq.(\ref{mass2}) there are no restrictions on masses of the three particles. 
For the one-electron $A^{+} B^{+} e^{-}$ and $A^{+} A^{+} e^{-}$ ions we can choose in 
Eq.(\ref{mass2}) $m_3 = m_e, m_1 = M_1$ and $m_2 = M_2$, where $M_2 \ge M_1$. In this notation and 
in atomic units, where $\hbar = 1, m_e = 1, e = 1$, the formula, Eq.(\ref{mass1}), and expressions 
for the parameters $\alpha, \beta, \gamma$ and $D$, Eq.(\ref{mass1}), are simplified substantially:
\begin{eqnarray}
 E = \frac12 \; \alpha \; \Phi( \beta, \gamma) \; \; , \; \; \alpha = \frac{M_1 \; M_2 }{D} 
 \; \; , \; \; \beta = \frac{M_1 + M_2}{2 \; D}  \; \; , \; \; \gamma = \frac{M_2 - M_1}{2 
 \; D} \; \; , \; \label{mass3}   
\end{eqnarray}
where $D = M_1 M_2 + M_1 + M_2$ and $M_{2} \ge M_{1}$. The dimensionless parameter $\gamma$ 
describes nuclear antisymmetry in the three-body, one electron ion $A^{+} B^{+} e^{-}$ which 
contains two heavy atomic nuclei $A$ and $B$ with unit electrical charges. This is the reason why 
the parameter $\gamma$ is often called the parameter of antisymmetry. 

In this Section we want to study the function $\Phi( \beta, \gamma)$ which depends upon the two 
dimensionless variables $\beta$ and $\gamma$. First, consider the case of symmetric $A^{+} A^{+} 
e^{-}$ ions, in which $M_1 = M_2 \gg m_e = 1$. The total energies of such two-center ions are 
also represented by the formula, Eq.(\ref{mass3}):
\begin{eqnarray}
 E = \frac12 \; \alpha \; \Phi( \beta, 0) \; \; , \; \; \alpha = \frac{M^{2}}{D} 
 \; \; , \; \; \beta = \frac{M}{D} \; \; \; {\rm and} \; \; D = M^{2} + 2 \; M \; 
 \; , \; \label{mass33}   
\end{eqnarray} 
Originally, the three $\alpha, \beta$ and $\gamma$ variables have been introduced to analyze 
various three-body systems with three comparable masses $m_1, m_2$ and $m_3$. However, for the  
adiabatic two-center ions with one electron and two large nuclear masses these variables are 
not convenient. To deal with the two-center ions we chose the $\alpha, \beta, \gamma$ and $D$
variables in the following way: $\alpha = 1, \beta = \frac12 \; \Bigl( \frac{M_1 +  M_2}{D} 
\Bigr) \Rightarrow \frac{1}{2 M_1} + \frac{1}{2 M_2}, \gamma = \frac12 \; \Bigl( \frac{M_2 
- M_1}{D} \Bigr) \Rightarrow \frac{1}{M_1} - \frac{1}{M_2}$, $D = M_1 M_2$ where $M_1 \le M_2$. 
For symmetric $A^{+} A^{+} e^{-}$ ions the parameter $\beta$ equals $\frac{1}{M}$, while the 
second parameter equals zero, i.e., $\gamma = 0$. In this case the formula, Eq.(\ref{mass33}), 
is written in the form  
\begin{eqnarray}
 E(\beta) \; = \; \Phi( \beta, 0) \; = \; A_0 + \sum^{K_{max}}_{k = 1} \; A_k \; \beta^{k} 
 = \; A_0 + \sum^{K_{max}}_{k = 1} \; \frac{A_k}{M^{k}} \; \; \; , \; \; \; \label{mass333}   
\end{eqnarray} 
where $k = 1, 2, \ldots, K_{max} \;$ are the positive integer numbers. The power series, 
Eq.(\ref{mass333}), which contains only integer powers of inverse nuclear mass $M$, is accurate, 
fast convergent and quite effective in actual applications. In reality, there are some obvious 
restrictions on applications of this formula to actual systems. These restrictions are described 
by the following theorem. If the considered two-center symmetric $A^{+} A^{+} e^{-}$ ions belong 
to the mass interval $[ M_1, M_2 ]$ which does not contain any singular mass-point, then the 
total energies of any systems from this mass interval are perfectly represented and can be 
approximated to very good numerical accuracy by the formula, Eq.(\ref{mass333}). However, if the 
maximal mass ($M_2$) equals infinity, then the formula, Eq.(\ref{mass333}), must be replaced by 
a different formula  
\begin{eqnarray}
 E(\beta) \; = \; \Phi( \beta, 0) \; =  \; A_0 + \sum^{K_{max}}_{k = 1} \; A_k \; \beta^{q} 
 \; = \; A_0 + \sum^{K_{max}}_{k = 1} \; \frac{A_k}{M^{\Bigl(\frac{k}{\sigma}\Bigr)}} \; \; 
 \; , \; \; \; \label{mass334}   
\end{eqnarray} 
where $q = \frac{k}{\sigma}$ is a rational number, while $\sigma$ is a symmetry number for the 
symmetric $A^{+} A^{+} e^{-}$ ion which correspond to the mass-singular boundary point. The 
series, which consist of rational powers of the mass ratio, are the so-called Puiseaux series 
\cite{Puis1} and \cite{Puis2}. Modern theory of the Puiseaux series and their applications to 
various physical problems are well described in Kato's book \cite{Kato} (see, also references 
therein). Our theorem mentioned above directly follows the results derived and mentioned in 
\cite{Kato}. 

For three-body systems with the unit electric charges there is only one mass-singular system 
and it coincides with the truly adiabatic ${}^{\infty}$H$^{+}_{2}$ ion. Formally, this 
two-center one-electron ion is out of the three-particle world. Indeed, the 
${}^{\infty}$H$^{+}_{2}$ ion has an additional, or `hidden' dynamical symmetry (see, e.g, 
\cite{Fro2025} and references therein) in comparison with the usual $A^{+} A^{+} e^{-}$ and 
$A^{+} B^{+} e^{-}$ ions, where all nuclear masses are finite. In fact, for each eigenstate 
in the ${}^{\infty}$H$^{+}_{2}$ ion the four following operators: $\pm \; \hat{L}_{z}$ and 
$\hat{\Lambda}_{\pm} = \hat{L}^{2} \; + \; \sqrt{- 2 E(Q_A \pm Q_B)} \; R \; \hat{A}_{z}$ are 
also conserve. Here $R$ is the internuclear distance, while $\hat{L}_{z}$ and $\hat{A}_{z}$ 
are $z-$components of the electron's angular momenta $\hat{\bf L}$ and Runge - Lenz vector 
$\hat{\bf A}$ \cite{Fro2025}, respectively. Also, the notation $E(Q_A \pm Q_B) = - 
\frac{(Q_A \pm Q_B)^{2}}{2 n^{2}_{\pm}} $ is the energy of the one-electron hydrogen atom 
with the nuclear charge $Q_A + Q_B (= 2)$ and $Q_A - Q_B (= 0)$, respectively \cite{Fro2025}. 
In our case these electrical charges are + 2 and 0, respectively. The notations $n_{\pm}$ 
stand for the corresponding principal quantum numbers. For $Q_A - Q_B = 0$ the operator 
$\hat{\Lambda}_{-}$ simply coincides with the square of the total angular momentum 
$\hat{L}^{2}$ (for more details, see, e.g., \cite{Edm}).   

Briefly, this means that for each finite-norm eigenfunction $\Psi$, where $\hat{H} \Psi = E 
\Psi$, where $\hat{H}$ is the Hamiltonian defined above, the corresponding expectation 
values are conserved during time-evolution of this system. The existence of four additional 
conserving operators immediately gives us $\sigma = 4$ in Eq.(\ref{mass334}) for the 
${}^{\infty}$H$^{+}_{2}$ ion. For usual $A^{+} A^{+} e^{-}$ and $A^{+} B^{+} e^{-}$ ions 
with the finite nuclear masses these operators do not conserve and $\sigma = 0$. The theorem 
mentioned above means that, if the ${}^{\infty}$H$^{+}_{2}$ ion is included in the area of 
mass variation, then the $E(\beta)$ function must be represented by the Puiseaux series, 
Eq.(\ref{mass334}), i.e., the series which contain the rational powers of mass ratio 
$\Bigl(\frac{M}{m_e}\Bigr)$, or inverse mass ratio $\Bigl(\frac{m_e}{M}\Bigr)$ and the 
symmetry number $\sigma$ equals 4. In these cases the powers $q$ in Eq.(\ref{mass334}) equal 
$q = \frac{k}{4}$ (see above). The fact that $\sigma = 4$ was first proven in \cite{BO} and 
in other our earlier papers (see, e.g., \cite{Fro2018A} - \cite{Fro2021}) by using some 
different approaches, including one pure numerical method.   

In \cite{Fro2018A} - \cite{Fro2021} we have found a simple criterion which indicates when 
it is possible to apply the formula, Eq.(\ref{mass333}), or we have to replace this simple 
formula with another more complicated formula, Eq.(\ref{mass334}). In actual applications 
such a criterion simply means that if the truly adiabatic ${}^{\infty}$H$^{+}_{2}$ ion and 
close three-body systems are not included in the original set of our `fulcrum' or `anchor' 
systems, then the regular power-type formula, Eq.(\ref{mass333}), must be used. For 
instance, if you consider the $A^{+} B^{+} e^{-}$ ions with relatively small nuclear masses,  
e.g., $\max (\frac{M_A}{m_e,}, \frac{M_B}{m_e}) \le \Bigl(\frac{M}{m_e}\Bigr) \le 500,000$,  
then the regular power-type formula, Eq.(\ref{mass333}), is an ideal choice for the 
mass-interpolation formula. In the opposite case, i.e., when the minimal mass ratio in our 
selected `anchor' systems is very large, e.g., $\Bigl(\frac{M}{m_e}\Bigr) \ge$ 5,000,000, 
the mass-interpolation formula must be constructed in the form of Puiseaux series, 
Eq.(\ref{mass334}), which contain the rational powers of mass ratios.  

Now, consider the case of non-symmetric $A^{+} B^{+} e^{-}$ ions. For such systems we have an 
additional complication, since the parameter of anti-symmetry $\gamma$ is non-zero and we have 
to come up with some reliable and accurate `power-type' series upon $\gamma$ for the total 
energies of non-symmetric $A^{+} B^{+} e^{-}$ ions. This problem has never been considered in 
earlier studies. Therefore, for such ions we have to develop some original procedure in which   
all our results from Table I must be used. It is clear that such a procedure can be based on 
the least squares method, since the number of linear coefficients which must be determined is 
much smaller than the number of equations which we need to solve (or system energies shown in 
Table I). 

\section{Partial adiabatic limits in non-symmetric diatomic ions}

By investigating a number of different sequences of the non-symmetric one-electron $A^{+} B^{+} 
e^{-}$ ions we have discovered an interesting phenomenon, which is called the partial adiabatic 
limit. To understand this phenomenon let us consider the consequence of $A^{+} B^{+} e^{-}$ ions 
in which the mass of nucleus $A$ is fixed, e.g., $M_A$ = 2000 $m_e$, while the mass of second 
heavier) nucleus $B$ increases from 2000 $m_e$ to the infinity. Suppose, we want to predict the 
general behavior of the bound state energies in such a series (or sequence) of non-symmetric 
$A^{+} B^{+} e^{-}$ ions. First of all, it is clear $a$ $priori$ that such an energy limit at 
$M_B \rightarrow \infty$ will be finite and its numerical value is a uniform function of the 
inverse mass of another heavy nuclei $M_{A}$, or the dimensionless ratio $\frac{M_{A}}{m_{e}}$. 
For the total energies of non-symmetric one-electron $A^{+} B^{+} e^{-}$ ions we can write the 
following general formula: 
\begin{eqnarray} 
   E = \sum_{k=0} C_k \; \Bigl(\frac{M_{A}}{M_{B}}\Bigr)^{k} = C_0 + \sum_{k=1} C_k \; 
   \Bigl(\frac{M_{A}}{M_{B}}\Bigr)^{k} \; \; , \; \; \label{Parlim1}
\end{eqnarray} 
where $C_k$ ($k$ = 0, 1, $\ldots$) are the numerical coefficients and $\frac{M_{A}}{M_{B}} 
(< 1)$ can be considered as the `small' parameter in this formula in that case when $M_B 
\rightarrow +\infty$. By using our results from Table I we evaluated the partial adiabatic 
limit for the $A^{+} B^{+} e^{-}$ ions in the $(2000)^{+} B^{+} e^{-}$ series as $\approx$ 
-0.59887(2) $a.u.$ For the $(3000)^{+} B^{+} e^{-}$ series this limit equals $\approx$ 
-0.59964(2) $a.u.$, while for the similar $(4000)^{+} B^{+} e^{-}$ series such a limit is 
$\approx$ -0.60005(2) $a.u.$ It is clear that the numerical values of these partial 
adiabatic limits increase when the mass $M_{A}$ grows. In fact, they converge to some finite 
value which is called the absolute adiabatic limit of the total energies for the one-electron 
$A^{+} B^{+} e^{-}$ ions with unit electrical charges.It is easy to predict that this absolute 
adiabatic limit equals -0.60263421449494646150905 $a.u.$, which coincides with the total 
energy of the ground $1 s \sigma$-state in the truly adiabatic ${}^{\infty}$H$^{+}_{2}$ ion. 
The problem of the partial adiabatic mass-limits has never been considered in earlier studies. 

Right now, based only on our data presented in this study it is hard to evaluate such partial 
mass limits for all possible nuclear masses $M_A$ to high accuracy. However, by performing 
numerical calculations for larger numbers of similar $A^{+} B^{+} e^{-}$ systems in order to 
extend our current version of Table I we can also try to increase the overall accuracy of our 
evaluations of the partial mass limits for some sequences of non-symmetric one-electron 
$A^{+} B^{+} e^{-}$ ions. 

\section{Bound state spectra in one-electron molecular ions} 

Another interesting problem which we want to consider is directly related with the bound 
state spectra in the one-electron molecular $A^{+} A^{+} e^{-}$ and $A^{+} B^{+} e^{-}$ 
ions. As follows from the results numerical calculations of similar ions the total number 
of bound states in such ions rapidly grows when the both nuclear masses in the $A^{+} 
B^{+} e^{-}$ ions increase. For instance, in the series of `muonic molecular' ions $p p 
\mu, p d \mu, p t \mu, d d \mu, d t \mu$ and $t t \mu$ the number of bound states 
increases from two (in all  ions with protium nuclei) to five (in all ions which contain 
deuterium) and even six (in the $t t \mu$ ion). In the one-electron H$^{+}_{2}$, HD$^{+}$, 
HT$^{+}$, D$^{+}_2$, DT$^{+}$, T$^{+}_2$ and other similar ions the total numbers of 
bound states equal many dozens and even hundreds. In reality, the energies of these states 
converge to the dissociation threshold $E_{tr}$, which describes the actual two-body 
dissociation of the $A^{+} B^{+} e^{-}$ ion, i.e., $A^{+} B^{+} e^{-} =  A^{+} \; + \; 
B^{+} e^{-}$. For heavy and very heavy nucleus $B$ this means that we have to add 
$\approx$ 0.5 $a.u.$ to the negative total energy of each bound state in the $A^{+} B^{+} 
e^{-}$ ion. The arising new `binding' energies will converge to the to the new 
dissociation threshold $E_{tr} = 0$. This point is a limiting point for eigenvalues of 
the Hamiltonian, Eq.(\ref{Hamilt}), of the $A^{+} B^{+} e^{-}$ ion. Moreover, it is easy 
to show that this limiting point is unique, and the multiplicities of other eigenvalues 
of this Hamiltonian, which are different from $E_{tr} = 0$, are always finite. 

Formally, for any atomic (or one-center) system one finds the same situation - the binding 
energies converge to the corresponding ionization threshold and this point is always we 
need to define the Hamiltonian of the discrete spectra $\hat{H}_{-}$ \cite{Fro1999}. In 
general, our original Hamiltonian, Eq.(\ref{Hamilt}), is a self-adjoint operator and 
contains the both discrete and continuous spectra. This Hamiltonian $\hat{H}$ determines 
its unique spectral function $\hat{E}(\lambda)$ and can be written as the following 
spectral integral 
\begin{eqnarray} 
  \hat{H} = \int_{-\infty}^{+\infty} \lambda \; d\hat{E}({\lambda}) = \sum_{i=1}^{N} 
  \lambda_{i} \; \hat{E}_{i} + \int_{\lambda_{tr} - \varepsilon}^{+\infty} \lambda \; 
  d\hat{E}({\lambda}) \; \; , \; \; \label{SpectrH}
\end{eqnarray} 
where $N$ is a non-negative integer number, which can also be equal zero, or infinity. 
Below our main interest is related to the bound state spectrum, or `discrete spectrum', 
for short. Now, we can introduce the Hamiltonian of the discrete spectrum $\hat{H}_{-}$ 
which is defined as follows 
\begin{eqnarray} 
  \hat{H}_{-} = \sum_{i=1}^{N} \lambda_{i} \; \hat{E}_{i} + \int_{\lambda_{tr} - 
  \varepsilon}^{0} \lambda \; d\hat{E}({\lambda}) = \sum_{i=1}^{N} \lambda_{i} \; 
  \hat{E}_{i} + \int_{- \varepsilon}^{0} \lambda \; d\hat{E}({\lambda}) \; \; , 
  \; \; \label{SpectrH-}
\end{eqnarray} 
where without loss of generality we assumed that the actual threshold energy (or the 
corresponding eigenvalue $\lambda_{tr}$) equals zero identically. It is clear that 
this approach works for atoms, ions and molecular ions. Furthermore, as directly 
follows from Eq.(\ref{SpectrH-}) the Hamiltonian of the discrete spectra of the 
two-center $A^{+} B^{+} e^{-}$ ion and similar atomic Hamiltonians are the completely 
continuous operators. However, in the first case we are dealing with the nuclear (or 
kernel), completely continuous operator, while the Hamiltonians of the discrete 
spectra of atoms (and ions) belongs to the Hilbert-Schmidt class. The fundamental 
difference between these two classes of operators directly follows from their 
definitions and their properties (see, e.g., \cite{Maurin}). Let us consider the 
following spectral sum of eigenvalues $\lambda_{k}$, which must be taken with their 
finite multiplicities \cite{XXX}, which are designated here as ${\rm dim} \{ 
\phi(\lambda_k) \}$. This sum is written in the form: 
\begin{eqnarray} 
  S_{p} = \sum^{K}_{k=1} \mid \lambda_{k} \mid^{p} \; {\rm dim} \{ \phi(\lambda_k) \} 
  = \sum^{K}_{k=1} \mid \lambda_{k} \mid^{p} \; {\rm dim} \{ {\cal H}_{\lambda_k} \} 
  \; \; , \; \; \label{SP} 
\end{eqnarray} 
where the number $K$ can be finite, or infinite. The sum $S_{p}$ defined in Eq.(\ref{SP}) 
as a great value for classification of completely continuous operators. Now, if the sum 
$S_{p}$ in Eq.(\ref{SP}) converges for $p = 2$ (but diverges for $p = 1$), then the 
operator $\hat{H}_{-}$ is a Hilbert-Schmidt operator. This case corresponds to the bound 
state spectra of atoms and ions. In particular, for the hydrogenic atom(s) and 
hydrogen-like ions, where $\lambda_n = -\frac{Q^{2} e^{2} m_e}{2 n^{2} \hbar^{2}}$, one 
obtains 
\begin{eqnarray} 
  S_{2} = \frac{Q^{4} e^{8} m^{2}_e}{4 \hbar^{4}} \; \Bigl\{ \sum^{\infty}_{n = 1} 
  \frac{1}{n^4} \sum^{n - 1}_{\ell = 0} ( 2 \ell + 1 ) \Bigr\} = \frac{Q^{4} Ry^{2}}{4} 
  \Bigl( \frac{2 \pi^{2} \mid B_2 \mid }{2} \Bigr) = \frac{Q^{4} \pi^{2} Ry^{2}}{24} 
  \nonumber  
\end{eqnarray} 
where $Ry = \frac{e^{2} m_e}{\hbar^{2}}$ is the atomic Rydberg constant, $B_2 = - 
\frac16$ is the second Bernoulli number, while $Q$ is the nuclear charge of the central 
atomic nucleus. In atomic units, where $Ry = 1$, the last sum equals $S_{2} = 
\frac{Q^{4} \pi^{2}}{24}$ (see, also \cite{Fro2018}). In our earlier studies this finite 
sum $S_2$ was called the Hilbert-Schmidt sum. It is clear that for actual few- and 
many-electron atoms and ions the Hilbert-Schmidt sum $S_2$ must also include terms which 
describe the electron-electron correlations. Furthermore, for similar systems we have to 
re-define the actual threshold energies (or zero-energies) for the bound state spectrum. 
Unfortunately, here we cannot discuss (even briefly) this interesting problem. Note that 
for any neutral atom and positively charged ion the sum $S_2$ is finite, but similar 
`nuclear' sum $S_{1}$ always diverges.  

In general, if the sum $S_{p}$ in Eq.(\ref{SP}) converges for $p = 1$ (but diverges for $p 
= 0$), then the operator $\hat{H}_{-}$ is a nuclear (or kernel) completely continuous 
operator. This case corresponds to the pure adiabatic ${}^{\infty}$H$^{+}_{2}$ ion and 
other similar two-center atomic ions. In reality, the bound state spectra of the pure 
adiabatic ${}^{\infty}$H$^{+}_{2}$ ion for each $R \ne 0$, where $R$ is the internuclear 
distance, is represented as the spectrum of a nuclear completely continuous operator. This 
important result explicitly shows the main difference between the bound state spectra in 
atoms and similar spectra of the two-center molecular ions. For the first time such a 
difference was described in \cite{Fro1999}. Note that the bound state spectrum of an 
arbitrary three-body system with the unit atomic charges always contains a finite number 
of bound states, i.e., it is a finite spectrum \cite{Fro1999}. This includes the bound 
state spectra of molecular H$^{+}_{2}$, HD$^{+}$, HT$^{+}$, etc, ions which have dozens 
and hundreds of bound states. However, the actual nuclear (or kernel) spectrum of bound 
states arise only in the pure adiabatic limit, i.e., for the ${}^{\infty}$H$^{+}_{2}$ ion 
at any finite internuclear distance $R$. However, in the limit of united atom, when $R = 
0$ we arrive to the Hilbert-Schmidt bound-state spectrum of the one-electron 
${}^{\infty}$He$^{+}$ ion. Similar situation arise for two separated `atoms', when $R = 
+ \infty$. In all other cases we are dealing with the nuclear (or kernel) spectrum of bound 
states in the ${}^{\infty}$H$^{+}_{2}$ ion. 

To conclude this Section we want to note here that first analytical and numerical investigations 
of the bound state spectra in the ${}^{\infty}$H$^{+}_{2}$ ion were performed by E. Teller 
\cite{Teller}, Hylleraas \cite{Hyl1931} and Jaffe \cite{Jaffe}. Eyring always noticed (see, e.g., 
\cite{Eyr}) that for molecules and molecular ions the three-body ${}^{\infty}$H$^{+}_{2}$ ion 
plays the same fundamental role that the hydrogen atom plays for all atoms

\section{Discussion and Conclusion} 

We investigated highly accurate calculations of bound states in the one-electron diatomic (or 
two-center) $A^{+} A^{+} e^{-}$ and $A^{+} B^{+} e^{-}$ ions. In such ions the both nuclear 
masses $M_A$ and $M_B$ are significantly larger than the electron mass $m_e$. In this study 
we restricted by the analysis of the bound (ground) $1 \sigma-$states in the three-body ions 
with the unit electrical charges. However, our general method, procedure and even conclusions 
can be applied to the excited bound states in various one-electron diatomic ions with 
arbitrary electrical charges. At this moment is absolutely clear that the adiabatic (or 
two-center) Born-Oppenheimer approximation do not play any role in highly accurate bound state 
calculations of the one-electron diatomic (or two-center) $A^{+} A^{+} e^{-}$ and $A^{+} B^{+} 
e^{-}$ ions. Indeed, the current accuracy of our method based on the modern three-body approach 
is significantly higher than the old-fashion Born-Oppenheimer approximation can ever provide. 

On the other hand, our method provides only minimal (or extremal) numerical solution of the 
Schr\"{o}dinger equation for one-electron diatomic $A^{+} A^{+} e^{-}$ and $A^{+} B^{+} e^{-}$ 
ions and we cannot plot any of old-fashion pictures known from earlier studies as `molecular 
(energy) terms'. Nevertheless, our current method is a very effective, relatively simple and 
highly accurate tool which allows one to determined and investigate the bound states and their 
properties in various one-electron diatomic ions. In fact, at this moment we still have quite 
a few unsolved problems for similar three-body systems and in this study we investigated some 
of such problems. 

In particular, we developed and tested a few mass-interpolation formulas constructed for the 
bound state energies in one-electron diatomic ions. It is shown that for the light $A^{+} A^{+} 
e^{-}$  and $A^{+} B^{+} e^{-}$ ions it is always possible to apply the regular power-type mass 
series. These mass-series are relatively simple in applications, but they provide a sufficient 
numerical accuracy in actual applications. In this study we developed some accurate 
mass-interpolation formulas for the total energies $E(\beta)$ of all symmetric $A^{+} A^{+} 
e^{-}$ ions which contain two heavy atomic nuclei and one bound electron. It is shown that for 
actual $A^{+} A^{+} e^{-}$ ions, where $M_A \le 25,000$ $m_e$, we need to apply our formula, 
Eq.(\ref{mass333}) with integer powers $k$. This simple formula has a number of advantages in 
comparison with the more complex, competing formula, Eq.(\ref{mass334}), which is based on the 
Puiseaux series. For the symmetric two-center $A^{+} A^{+} e^{-}$ ions a number of accurate 
mass-interpolation formulas have been derived in earlier studies (see, e.g., \cite{Fro2018A} - 
\cite{Fro2021}). However, for the non-symmetric two-center $A^{+} B^{+} e^{-}$ ions the explicit 
construction of highly accurate mass-interpolation formulas is much more difficult process which 
includes a number of non-trivial steps. Nevertheless, in this study we could derive and test our 
accurate mass-interpolation formula which represents the total energies $E(\beta,\gamma)$ of the 
non-symmetric $A^{+} B^{+} e^{-}$ ions. 

In physics of three-particle systems, there has always been interest to understand the origin of 
atomic `adiabaticity'. In other words, there was interest in understanding how, from a fairly 
general three-particle (Coulomb) system, a system (ion) with the two heavy centers arises during 
smooth changes of basic physical parameters. In particular, Hans Bethe was strongly interested 
in this problem and originally he wanted to include its discussion in his famous book \cite{BS}. 
Since 1990's a large number of research of the two-center atomic ions have been performed with 
the use of methods which are more traditional for the general three-body atoms and ions (see, 
e.g., \cite{BaSh} - \cite{HPRA1}). Currently, the number of publications in this area is growing 
very rapidly and it is already difficult to mention all of them in a single article. Here we want 
to mention only the review of atomic `adiabaticity' written by B. Sutcliffe in \cite{BT} and the 
textbook \cite{EluKri}, where the regular and two-center three-body ions were considered on equal 
basis.  

In conclusion, we want to note that recent extensive research of the two-center, one-electron 
$A^{+} A^{+} e^{-}$ and $A^{+} B^{+} e^{-}$ ions substantially increased our understanding of 
bound states in such diatomic systems. Many long-standing questions, which were existed for 
such systems and ions, have been answered completely and accurately. On the other hand, it is 
clear that further investigations are needed in this area of research. 

\begin{center}
   {\bf Acknowledgments:} 
\end{center} 

Personally, for me it was a great pleasure to take a leading role in the solution of this 
`central' problem which was known for years in the three-body physics. However, these my 
project could not be finished without real and valuable help from Yurii N. Demkov 
(Saint Petersburg, Russia) and David H. Bailey (Berkeley, California, USA). Also, I wish 
to thank Prof. G.W.F. Drake (University of Windsor, Windsor, Ontario) who has provided 
sufficient computational resources, which were critical to finish this project. \\


\begin{center}
   {\bf Declarations} 
\end{center} 

\noindent
{\bf Author contributions}: Alexei M. Frolov contributed to the 
conceptualization (equal); formal analysis (equal); investigation 
(equal); writing—original draft (equal); and writing—review and 
editing (equal). \\

\noindent
{\bf Data Availability Statement}: This manuscript has no associated 
data or the data will not be deposited. [Author's comment: All data 
generated or analyzed during this study are included in this published 
article and cited references]. \\

\noindent
{\bf Funding Statement}: I received no funding to write/supervise 
this manuscript and conduct all related research. \\ 

\noindent
{\bf Ethical Statement}: Currently, I have no `conflicts of interests' 
which can be associated with this manuscript. \\ 


\begin{center}
{\bf Appendix A.} \\
{\bf Relations between three-body perimetric and ellipsoidal coordinates}
\end{center}

In this Appendix we discuss an interesting question about properties of the perimetric three-body 
coordinates and derive the explicit relation between these coordinates and ellipsoidal coordinates. 
The ellipsoidal coordinates can be introduced in any one-electron three-body system with the two 
heavy nuclei (see, e.g., \cite{LLQ}, \cite{MQW} and \cite{AS}). Here we want to obtain the relation 
between the ellipsoidal and perimetric coordinates written in the form of exact equations. As 
mentioned in the main text in any three-body problem it is convenient to use three scalar relative 
coordinates $r_{32}, r_{31}$ and $r_{21}$ which coincide with the corresponding ribs of the triangle 
of particles (1,2,3). These relative coordinates are defined as follows: $r_{ij} = \mid {\bf r}_{i} 
- {\bf r}_{j} \mid$, where $(i,j,k)$ = (1,2,3), while ${\bf r}_{i}, {\bf r}_{j}$ and ${\bf r}_{k}$ 
are the Cartesian coordinates of three particles. As follows from this definition we always have the 
equality $r_{ji} = r_{ij}$ for any two particles $i$ and $j$. Now we can introduce three scalar 
perimetric coordinates $u_1, u_2, u_3$ which are simply (even linearly) related to the relative 
coordinates and vice versa:  
\begin{eqnarray}
  & & u_1 = \frac12 (r_{21} + r_{31} - r_{32}) \; \; \; , \; \; \; r_{32} = u_2 
  + u_3 \; \; , \; \nonumber \\
  & & u_2 = \frac12 (r_{21} + r_{32} - r_{31}) \; \; \; , \; \; \; r_{31} = u_1 
  + u_3 \; \; , \; \label{ucoord} \\
  & & u_3 = \frac12 (r_{31} + r_{32} - r_{21}) \; \; \; , \; \; \; r_{21} = u_1 
  + u_2 \; \; , \; \nonumber
\end{eqnarray}
where $r_{ij} = r_{ji}$ are the relative coordinates defined above. The properties of these three-body 
perimetric coordinates $u_1, u_2$ and  $u_3$ are unique. First, these three coordinates are independent 
of each other. Second, each of these three coordinates is always non-negative. Third, each of these 
coordinates varies between zero and infinity, i.e., $0 \le u_{i} < + \infty$ for $i$ = 1, 2, 3. Unique 
combination of these three properties means that the original three-dimensional space of internal 
relative coordinates $R_{123} = r_{32} \biguplus r_{31} \biguplus r_{21}$ essentially splits into a 
direct product of three one-dimensional spaces, i.e., $R_{123} = U_1 \otimes U_2 \otimes U_3$, where 
$u_{i} \subset U_{i}$ for $i$ = 1, 2, 3, while the $U_i$ and $U_j$ spaces are orthogonal to each other. 
Furthermore, in many cases the arising three-dimensional integrals in relative coordinates are 
represented as finite sums of products of three one-dimensional integrals in perimetric coordinates. 
In addition to this, many of the arising one-dimensional integrals are reduced to the well known 
analytical expressions. Note also that in perimetric coordinates the area of any triangle is written 
in the form $S_{tr} = \sqrt{u_1 u_2 u_3 (u_1 + u_2 + u_3)}$ which explicitly shows symmetry between 
all vertexes of this triangle and between all of its edges (or ribs). It is easy to see that this 
formula retains its form, or it is invariant, under any permutation of the perimetric coordinates. 
Briefly, it is possible to say that the perimetric coordinates $u_1, u_2$ and $u_3$ are the natural 
three-body coordinates and they most convenient coordinates for applications to various three-body 
systems. 

Note also that the three-body perimetric (or triangle) coordinates were known to Archimedes $\approx$ 
2250 years ago and Hero of Alexandria $\approx$ 2000 years ago. A great advantage of these coordinates 
for the ancient Greeks was obvious, since they allow to determine the areas of arbitrary triangles by 
using only the lengths of their sides and ignoring any angles. This was a major achievement for ancient 
geometers, since they could determine to very good accuracy the area of any surface figure that can be 
divided into a number of triangles. On the other hand, it also gave some advantages to real estate 
agents of that time to carry out their traditional brainwashing. In modern few-body physics they have 
been introduced by C.L. Pekeris in 1958. Note that our definition of these perimetric coordinates, 
Eq.(\ref{ucoord}), slightly differs from their definition used by Pekeris. 

Now, let us derive the relations between the three-body perimetric coordinates and ellipsoidal 
coordinates $\xi, \eta,\phi$ which are often used for investigation of some Coulomb two-center 
systems with the two heavy atomic nuclei (molecular ions). Below, we shall assume that the 
inter-nuclear distance $r_{21} = R$ is a constant of motion, i.e., it does not change during 
the electron's motion. This corresponds to the one-electron, two-center ${}^{\infty}$H$^{+}_{2}$ 
ion which has the two infinitely heavy nuclei. In this case, the system Eq.(\ref{ucoord}) is 
reduced to the form 
\begin{eqnarray}
  u_1 = \frac12 \; R \; ( 1 - \eta ) \; \; , \; \; u_2 = \frac12 \; R \; ( 1 + \eta ) 
  \: \: , \; \; u_3 = \frac12 \; R \; ( \xi - 1 ) \; \; , \; \label{ucoordA} 
\end{eqnarray}
where $r_{21} = R$ and $\xi = \frac{r_{32} + r_{31}}{R}$ and $\eta = \frac{r_{32} - r_{31}}{R}$ 
are the two ellipsoidal coordinates which are traditionally defined in any one-electron problem 
with the two infinitely heavy Coulomb centers (see, e.g., \cite{LLQ}). From this system of 
equations one obtains the following constraint: $\frac{u_1}{R} + \frac{u_2}{R} = 1$, or $u_1 
+ u_2 = R$, for the first two perimetric coordinates $u_1$ and $u_2 $. This means that in any 
one-electron two-center system/ion these two perimetric coordinates $u_1$ and $u_2$ depend upon 
each other and one of them can always be excluded. \\

\begin{center}
{\bf Appendix B.} \\
{\bf Completeness of the exponential basis functions for three-body systems}
\end{center}

In this Appendix we want to answer the principal question about completeness of the exponential 
sets of basis functions which are used in our trial wave functions in Eqs.(\ref{expc}) and 
(\ref{expc1}) above. This problem is of great interest in various applications of these both 
exponential variational expansions. First, consider the case when $L = 0$ and all angular 
functions equal unity. In this case, all angular factors ${\cal Y}_{LM}^{(\ell_{1},\ell_{2})} 
({\bf r}_{31}, {\bf r}_{32})$ in formulas, Eqs.(\ref{expc}) and (\ref{expc1}), equal unity 
identically. Second, the factor $\frac{1}{2} (1 + \kappa \hat{P}_{21})$ which provides the
proper permutations between two identical particles in the wave function is also replaced by 
the unity. After these transformations the two expansions, Eqs.(\ref{expc}) and (\ref{expc1}), 
are reduced to the forms:  
\begin{eqnarray}
 \Psi_{LM}( r_{21}, r_{31}, r_{32}) = \sum_{i=1}^{N} C_{i} \exp(-\alpha_{i} r_{32} 
 - \beta_{i} r_{31} - \gamma_{i} r_{21}) \; \; \; , \; \; \label{exprelA} 
\end{eqnarray}
and 
\begin{eqnarray}
 \Psi_{LM}( u_1, u_2, u_3) = \sum_{i=1}^{N} C_{i} \exp(-\alpha_{i} u_1 - \beta_{i} u_2 
 - \gamma_{i} u_3) \; \; \; . \; \; \label{expupmA} 
\end{eqnarray}
By using the linear relations between the relative $( r_{32}, r_{31}, r_{21})$ and perimetric 
$( u_1, u_2, u_3)$ coordinates (see, Appendix A) we can re-write Eq.(\ref{exprelA}) into a 
different (but equivalent) form  
\begin{eqnarray}
 \Psi_{LM}( u_3, u_2, u_1) = \sum_{i=1}^{N} C_{i} \exp( - [\alpha_{i} + \beta_{i} ] u_{3} 
 - [\alpha_{i} + \gamma_{i} ] u_{2} - [\beta_{i} + \gamma_{i} ] u_1 ) \; \; 
 \; . \; \; \label{exprel1} 
\end{eqnarray}
Now, it is easy to understand that the two last equations and there is an uniform 
correspondence between them: $\alpha_{i} \Leftrightarrow \beta_{i} + \gamma_{i} , \beta_{i} 
\Leftrightarrow \alpha_{i} + \gamma_{i}$ and $\gamma_{i} \Leftrightarrow \beta_{i} + 
\alpha_{i}$. In other words, the both Eqs.(\ref{expupmA}) and (\ref{exprelA}) are, in fact, 
the same exponential expansions written in perimetric coordinates. To simplify our analysis 
below we restrict ourselves to the case of Eq.(\ref{expupmA}) only. 

Let us consider the set of exponential basis functions from Eq.(\ref{expupmA}) with different 
non-linear parameters $\alpha_{i}, \beta_{i}, \gamma_{i}$ and indexes ($i$). These `pure' 
exponential functions, and all finite linear combinations of such exponential basis from 
Eq.(\ref{expupmA}) can be included in one family of exponential functions which is designated 
here as ${\cal E}$. Briefly, the family ${\cal E}$ is the vector subspace spanned by the 
`pure' exponential functions in three-body perimetric coordinates, or the span of `pure' 
exponential functions, for short. In general, if we choose any two exponential functions from 
this family, then their product also belongs to the same family ${\cal E}$. In other words, 
all exponential functions form the closed, commutative group under multiplication. For our 
present purposes it is important to know when the family ${\cal E}$ can generate a complete 
basis set of functions in the three-dimensional space of perimetric coordinates $U_{123} = 
U_1 \otimes U_2 \otimes U_3$. The answer to this question is simple in one-dimensional case: 
\textit{the system of exponential functions is complete, if a series of inverse powers of the 
`non-linear parameters', or $S_1 = \sum^{\infty}_{i=1} \frac{1}{\alpha_{i}}$, \; diverges}. 
For the exponential functions, Eq.(\ref{expupmA}), of three perimetric coordinates one finds 
the three following and different conditions which must be obeyed simultaneously, or in other 
words, if the three following series $S_1 = \sum^{\infty}_{i=1} \frac{1}{\alpha_{i}} \; , \; 
S_2 = \sum^{\infty}_{i=1} \frac{1}{\beta_{i}} \; , \; S_3 = \sum^{\infty}_{i=1} 
\frac{1}{\gamma_{i}}$ diverge, then the exponential functions of three perimetric coordinates 
form a complete system of functions in the three-dimensional space of perimetric coordinates 
$U_{123} = U_1 \otimes U_2 \otimes U_3$. 

From now on everywhere in this Section we shall assume that these three conditions are always 
obeyed for `our' functions from ${\cal E}$. Other exponential functions are not of interest 
for our current analysis. Now, we can say that the family ${\cal E}$ is a closed algebra, or 
algebra of complex exponential functions $f(u_1, u_2, u_3)$ which depend upon the three 
perimetric coordinates. In general, a family ${\cal A}$ of complex functions defined on a set 
$E$ is said to be an algebra if the three following conditions are obeyed: (1) $f + g \in 
{\cal A}$, (2) $f \; g \in {\cal A}$ for all $f \in {\cal A}, g \in {\cal A}$ and $c \; f 
\in {\cal A}$ for all complex constants $c$. This means that the algebra ${\cal A}$ is a 
closed structure under addition, multiplication and scalar multiplication. In particular, 
these three conditions obey for our family of functions ${\cal E}$, i.e., this family is the 
algebra of exponential functions. 

Let us investigate some additional and interesting properties of this algebra. First, based on the 
known properties of exponential functions, it is possible to show that, if we choose two arbitrary 
different points $(u^{(1)}_1, u^{(1)}_2, u^{(1)}_3)$ and $(u^{(2)}_1, u^{(2)}_2, u^{(2)}_3)$ in the 
three-dimensional $U_{123} = U_1 \otimes U_2 \otimes U_3$ space mentioned above (space of three 
perimetric coordinates), then it is always possible to find some exponential function $g$ from our 
algebra ${\cal E}$, for which $g(u^{(1)}_1, u^{(1)}_2, u^{(1)}_3) \ne g(u^{(2)}_1, u^{(2)}_2, 
u^{(2)}_3)$. Briefly, this means that the algebra of exponential functions ${\cal E}$ separates 
points on $U_{123}$. Another property of the algebra ${\cal E}$ is equally important for our 
purposes in this study. Indeed, by using the known properties of usual exponents one can prove the 
following statement. For an arbitrary point $( u^{(0)}_1, u^{(0)}_2, u^{(0)}_3)$ in the $U_{123}$ 
space it is possible to find the function $f$ such that $f( u^{(0)}_1, u^{(0)}_2, u^{(0)}_3) \ne 
0$. This means that our algebra ${\cal E}$ vanishes at no point of the $U_{123}$ space. The unique 
combination of these two properties, i.e., separation of any two different points in the $U_{123}$ 
space and no-vanishing nature of our algebra ${\cal E}$, allows us to prove the following statement 
which is known as the Stone-Weierstrass theorem. 

In order to make our current analysis completely rigorous, let consider the exponential functions 
of the perimetric coordinates defined only at the compact sets $K$ from the $U_{123}$ space (not 
in the whole $U_{123}$ space). For instance, we can define such a compact set $K$ as follows: 
\begin{eqnarray}
  0 \le u_{1} \le B \; \; , \; \; 0 \le u_{2} \le B \; \; , \; \; 0 \le u_{3} \le B \; \; , \; 
  \; \label{interval} 
\end{eqnarray}
where $B$ is a very large real number. In other words, this is a bounded interval in the 
three-dimensional space of perimetric coordinates $U_{123} = U_1 \otimes U_2 \otimes U_3$. Now, 
consider the algebra ${\cal E}$ of exponential functions of perimetric coordinates, 
Eq.(\ref{expupmA}), on this compact interval $K$. It is easy to see that this algebra ${\cal E}$ 
of exponential functions separates points on $K$ and vanishes at no point of $K$. Therefore, we 
can apply the well known Stone-Weierstrass theorem (see, e.g., \cite{Rudin}), which leads us to 
the conclusion that uniform closure of ${\cal E}$ consists of all continuous functions on $K$, 
where $K$ is an arbitrary compact subset from the $U_{123}$ space of perimetric coordinates $u_1, 
u_2, u_3$. Briefly, this means that we can approximate any arbitrary continuous functions on $K$ 
by using series of exponential  functions from ${\cal E}$. This also means that the mentioned 
algebra ${\cal E}$ of all exponential functions on $K$ can be used to approximate (to an 
arbitrary, in principle, precision) our unknown wave function(s) written in the perimetric 
coordinates.Another variant of the same Stone-Weierstrass theorem can be proven for the `all 
continuous real functions $f( u_1, u_2, u_3)$ on $K$'. However, for this we have to assume that 
the numerical constants $c$, which are mentioned in the point (3) above, are real (not complex!). 

Thus, we have shown that the system of our exponential functions written in three perimetric 
coordinates is complete in the $U_{123} = U_1 \otimes U_2 \otimes U_3$ space. Now, we need to 
generalize our approach and show the completeness of exponential functions for the rotationally 
excited states in three-body systems, or for bound states with $L \ge 1$. In reality, these 
rotationally excited states in three-body systems can be considered as we analyzed the bound 
$S(L = 0)-$states. However, we need to prove the following statement: the product of any two 
exponential functions each of each now contains an additional angular factor ${\cal 
Y}_{LM}^{(\ell_{1},\ell_{2})}({\bf r}_{31}, {\bf r}_{32})$ is another exponential function 
with the similar angular factor, or it is expressed as a linear combinations of a number of 
exponential functions with similar angular factors. This statement plays a fundamental role 
in the whole theory, since it allows us to operate with the closed algebra of exponential 
functions in the case of rotationally excited states. This means that we need to obtain and 
then apply the closed analytical expressions for the two following angular integrals: 
\begin{eqnarray}
  I^{(\ell_{1}, \ell_{2}, \ell_{3}, \ell_{4})}_{LM} = \oint \oint d{\bf r}_{31} 
  d{\bf r}_{31} {\cal Y}_{LM}^{(\ell_{1},\ell_{2})}({\bf r}_{31}, {\bf r}_{32}) 
  {\cal Y}_{LM}^{(\ell_{3},\ell_{4})}({\bf r}_{31}, {\bf r}_{32})  \nonumber 
\end{eqnarray} 
and 
\begin{eqnarray}
  J^{(\ell_{1}, \ell_{2}, \ell_{3}, \ell_{4}, \ell_{5}, \ell_{6})}_{LM} = \oint 
  \oint d{\bf r}_{31} d{\bf r}_{31} {\cal Y}_{LM}^{(\ell_{1},\ell_{2})}({\bf r}_{31}, 
  {\bf r}_{32}) {\cal Y}_{LM}^{(\ell_{3},\ell_{4})}({\bf r}_{31}, {\bf r}_{32})  
  {\cal Y}_{LM}^{(\ell_{5},\ell_{6})}({\bf r}_{31}, {\bf r}_{32}) \nonumber 
\end{eqnarray} 
In reality, the analytical formulas for these two integrals are known for quite some time 
(they can be found, e.g., in \cite{FroWa} -\cite{Efro}). Each of these integrals is 
represented as a function of pure radial variables: $r_{32}, r_{31}, r_{21}$ (relative 
coordinates), or $u_1, u_2, u_3$ (perimetric coordinates). Analytical formulas for 
calculations of these radial integrals were explicitly derived in \cite{FroWa}. Furthermore, 
our paper \cite{FroWa} contains an extensive discussion of separation of three-body angular 
and radial variables. By using these facts it is possible to prove that our exponential 
functions with the additional angular factors form the closed algebra (exactly as for the 
$L = 0$ states). In general, this algebra can be designated by the same letter ${\cal E}$ 
and this algebra is also defined in the $U_{123} = U_1 \otimes U_2 \otimes U_3$ space. Then 
we can introduce the compact set $K$ (exactly as we did above for the $L = 0$ states). This 
allows us to prove the completeness of the algebra of exponential functions for the 
rotationally excited states. A few small differences with the $L = 0$ bound states, which 
arise during this procedure, are not of great interest and can be ignored in this study.

\begin{table}[tbp]
   \caption{The total energies of the ground states (or $1 s \sigma-$states) of some two-center \linebreak
            $(A B e)^{+} (= A^{+} \; B^{+} \; e^{-})$ ions in atomic units$^{(a)}$.} 
     \begin{center}
     \scalebox{0.985}{%
     \begin{tabular}{| c | c | c | c |}
      \hline\hline
 $M_B$  & $M_A$ = 2000 $m_e$ & $M_A$ = 3000 $m_e$ & $M_A$ = 4000 $m_e$ \\ 
     \hline   
  2000 & -0.5973769059509827425552 & ------------------------- & ------------------------- \\ 
  3000 & -0.5978494961316409303705 & -0.5983689922369481787515 & ------------------------- \\ 
  4000 & -0.5981024820645748115284 & -0.5986518542394255525217 & -0.5989546148412066735702 \\   
  5000 & -0.5982606723291479842128 & -0.5988308496364925030227 & -0.5991479210144068354459 \\
  6000 & -0.5983691153825283662958 & -0.5989546489521144889137 & -0.5992825439433229232929 \\
  7000 & -0.5984481531694510339140 & -0.5990455008314326706421 & -0.5993818836319580817072 \\ 
  8000 & -0.5985083451551111894851 & -0.5991150698960675353278 & -0.5994582942964640573229 \\
  9000 & -0.5985557266650720714154 & -0.5991700781702723400787 & -0.5995189381980423619197 \\
 10000 & -0.5985940006640559267948 & -0.5992146785401468362773 & -0.5995682632734673405899 \\ 
    \hline\hline
 $M_B$  & $M_A$ = 5000 $m_e$ & $M_A$ = 6000 $m_e$ & $M_A$ = 7000 $m_e$ \\ 
     \hline\hline 
  5000 & -0.5993517730213303139692 & ------------------------- & ------------------------- \\ 
  6000 & -0.5994945257602948267560 & -0.5996436428556994395796 & ------------------------- \\ 
  7000 & -0.5996003395936599123159 & -0.5997545894332993136908 & -0.5998697195621938224580 \\
  8000 & -0.5996820352266948710325 & -0.5998405196552600348665 & -0.5999591324908004425252 \\
  9000 & -0.5997470789319417948657 & -0.5999091206070951317394 & -0.6000306818314740280666 \\
 10000 & -0.5998001260542417616059 & -0.5999652004287618243264 & -0.6000892922747042028525 \\
      \hline\hline
 $M_B$  & $M_A$ = 8000 $m_e$ & $M_A$ = 9000 $m_e$ & $M_A$ = 10000 $m_e$ \\ 
     \hline\hline       
  8000 & -0.6000514672386532876384 & ------------------------- & ------------------------- \\ 
  9000 & -0.6001255068559395206583 & -0.6002016815348805397392 & ------------------------- \\ 
 10000 & -0.6001862674000662812864 & -0.6002642951605840606940 & -0.6003285246477842788322 \\
     \hline\hline
  \end{tabular}}
  \end{center}
${}^{(a)}$The notation $p \; q \; e$ system (or $(p \; q \; e)^{+}$ ion) means the two-center 
(adiabatic) $A^{+} B^{+} e^{-}$ ion in which the mass $M_{A} = p \cdot 1000 \; m_{e}$, while 
the mass $M_{B} = q \cdot 1000 \; m_{e}$ (here for all systems we assume that $M_{A} \le 
M_{B}$) and $m_{e}$ is the electron mass, i.e, $m_{e} = 1$ in atomic units.
  \end{table}
%
\begin{table}[tbp]
   \caption{Convergence of the total energies of the ground states (or $1 s \sigma-$states) 
            for some two-center $(A B e)^{+}$ ions in atomic units$^{(a)}$. The notation 
            $N$ stands for the total number of basis functions (complex exponents) used.} 
     \begin{center}
     \scalebox{0.985}{%
     \begin{tabular}{| c | c | c |}
      \hline\hline
  $N$    &      $(5 \; 10 \; e)^{+}$       &     $(7 \; 7 \; e)^{+}$  \\  
     \hline
  4000 & -0.5998001260542417545630 & -0.5998697195621938224565 \\ 
  4200 & -0.5998001260542417594169 & -0.5998697195621938224576 \\ 
  4400 & -0.5998001260542417616059 & -0.5998697195621938224580 \\  
   \hline\hline
 $N$    &  $(5.5 \; 7.5 \; e)^{+}$  &  $(15 \; 21 \; e)^{+}$  \\  
     \hline
  4000 & -0.5997280085527341809044 & -0.60089818489805598287846 \\ 

  4200 & -0.5997280085527341849025 & -0.60089818489805629822970 \\ 

  4400 & -0.5997280085527341867046 & -0.60089818489805644147520 \\  
   \hline\hline
  \end{tabular}}
  \end{center}
${}^{(a)}$The notation $p \; q \; e$ system (or $(p \; q \; e)^{+}$ ion) means the 
two-center (adiabatic) $A^{+} B^{+} e^{-}$ ion in which the mass $M_{A} = p \cdot 
1000 \; m_{e}$, while the mass $M_{B} = q \cdot 1000 \; m_{e}$ and $m_{e}$ is the 
electron mass, i.e., $m_{e} = 1$ in atomic units.  
  \end{table}
\end{document}